\documentclass{article} 
\usepackage{graphicx}
\usepackage{caption}
\usepackage{subcaption}
\usepackage{iclr2026_conference,times}
\usepackage{amsmath}
\usepackage{amsfonts}

\usepackage{float}

\usepackage{hyperref}
\usepackage{url}
\usepackage{amsmath}

\def\({\left(}
\def\){\right)}
\def\[{\left[}
\def\]{\right]}
\def\<{\langle}
\def\>{\rangle}

\newcommand{\bmat}{\begin{bmatrix}}
\newcommand{\emat}{\end{bmatrix}}

\newcommand\half{{\ensuremath{\frac{1}{2}}}}
\newcommand\p{\ensuremath{\partial}}

\newcommand\vev[1]{{\ensuremath{\left\langle{#1}\right\rangle}}}

\newcommand{\be}{\begin{equation}}
\newcommand{\ee}{\end{equation}}
\newcommand{\bea}{\begin{eqnarray}}
\newcommand{\eea}{\end{eqnarray}}
\newcommand{\bwt}{\begin{widetext}}
\newcommand{\ewt}{\end{widetext}}

\newcommand{\bi}{\begin{itemize}}
\newcommand{\ei}{\end{itemize}}
\newcommand{\ben}{\begin{enumerate}}
\newcommand{\een}{\end{enumerate}}
\newcommand{\bca}{\begin{cases}}
\newcommand{\eca}{\end{cases}}
\newcommand{\bln}{\begin{align}}
\newcommand{\eln}{\end{align}}
\newcommand{\bst}{\begin{split}}
\newcommand{\est}{\end{split}}

\newcommand\al{{\alpha}}

\newcommand\lam{\lambda}

\newcommand\Om{\Omega}

\newcommand\ha{{\half}}

\def\le{\left}
\def\ri{\right}

\newcommand{\NI}[1]{\textcolor{blue}{\textsf{[NI: #1]}}}

\title{Information spreading in diffusion models from effective field theory}

\author{
Navonil Neogi\textsuperscript{1,2} \quad Nabil Iqbal\textsuperscript{1,3} \\
\textsuperscript{1}Centre for Particle Theory, Department of Mathematical Sciences, Durham University \\
\textsuperscript{2}Institute of Physics, University of Amsterdam \\
\textsuperscript{3}Amsterdam Machine Learning Lab, University of Amsterdam 
 \\
\texttt{navonil.neogi@durham.ac.uk, nabil.iqbal@durham.ac.uk}
}

\iclrfinalcopy 
\begin{document}

\maketitle

\begin{abstract}
We study score-matching diffusion models with a convolutional architecture. We argue that the inductive bias of locality means that the machinery of {\it effective field theory} from physics can be usefully applied to describe the denoising dynamics. We apply this formalism first to a simple toy example which permits an analytical description, and thereafter to MNIST, and show that in both cases, the mutual information between two points grows in a manner predicted by a simple effective field theory of Brownian motion.  

\end{abstract}

\section{Motivation}

One of the key insights of physics in the twentieth century is that of {\it effective theory}, i.e. the idea that one can understand systems scale-by-scale, and that the long-distance behavior of a given system can be understood {\it without} full knowledge of the microscopic behavior. In most conventional physical systems principles such as locality and symmetry allow one to formulate an effective description where the short-distance physics affects the long-distance dynamics only through a few relevant parameters. These ideas have found particularly powerful use when applied to quantum field theory \citep{Weinberg:1978kz}. 

In this work we turn the same philosophy to the understanding of {\it diffusion models} in deep learning. Such models are a workhorse of generative AI, and are used to generate samples from a data distribution by gradually constructing structure from a starting distribution of pure noise. In general the dynamics of such models is rather complicated; however, as we argue below, under certain circumstances principles of locality and symmetry can be used to derive a very simple effective field theory which governs the growth of mutual information at early times and long distances in the reverse process. We empirically test the predictions of our effective theory in very simple trained diffusion models on toy datasets. We find this to be an interesting and potentially useful application of the principles of effective theory in a new context. An attempt has been made to provide background material to make the paper more self-contained for members of both the physics and machine-learning communities. 


\subsection{Previous work}\textbf{Diffusion models} \citep{sohl2015deep,ho2020denoising} are a powerful and commonly used method in generative AI, achieving state of the art results across many modalities, including images \citep{dhariwal2021diffusion,rombach2022high}, text \citep{li2023diffusion}, and scientific applications such as construction of molecules  \citep{hoogeboom2022equivariant}. In this work we study these models in their score-matching incarnation \cite{song2020score}. The dynamical regimes we describe in the reverse process are similar to those detailed in \cite{NEURIPS2023_d0da30e3} \cite{Biroli_2024}, and in particular, we mainly study what appears to correspond to the 'Gaussian' regime in those works. However, our work considers spatial structure within the samples, and aims to better understand the 'Brownian motion' regime described there. We also briefly examine the 'memorisation' regime in \cite{Biroli_2024} - which we call the 'snapping' regime - where samples approach their nearest training equivalents. Previous work on understanding general principles underlying these models also includes \cite{cotler2023renormalizingdiffusionmodels} from a physical perspective.

In particular, an approach based on information theory has been applied to study the creation of structure in data within diffusion models. The so-called 'I-MMSE' relationship formulated the evolution of mutual information between successive Gaussian evolutions, first in \cite{guo2004mutualinfo} and later extended and applied to the pointwise case in \cite{venkat2012pointwiserelations, kong2023informationtheoretic}. A useful contribution of our work is to study spatial mutual information on a pixel level over the course of the reverse process in diffusion, using a specifically physics-based approach (EFT) for handling the spatial structure in the case of convolutional architecture.

The object we study in the regime near the beginning of the reverse process in our EFT description is also studied as the second-order term in the Jacobian in the recent work \cite{ambrogioni2026outofequilibriumphasetransitionsseed}. The same analytic formulation of the $\pm 1$ grid example is also discussed therein. A different perspective on the transitions between regimes where patterns are formed and patches snap is provided in terms of collapse of Fourier modes. It is our belief that these perspectives are connected and future work will explore these connections.

\textbf{Effective field theory (EFT)} is a classic approach in physics \citep{Weinberg:1978kz, Polchinski:1992ed} where constraints on the description of a system - in physics, typically locality, symmetry or causality - allow for the construction of simplified models that describe its dynamics. As we review, such simplified models can usually fruitfully be understood as understanding the system of interest at a particular length scale. We apply this approach to model the reverse process of a diffusion model, to understand how mutual information is created and spreads in the sample as it is gradually denoised.

\section{Theory}
\subsection{Diffusion models and what they do}

In recent years, diffusion models have become ubiquitous in machine learning to generate high-dimensional data, in particular images. Below we present a brief review for a physics audience. The reader who is already familiar with diffusion models and their score-matching formulation is encouraged to skip this short exposition and proceed to the effective field theory of the score function.

The problem of generative artificial intelligence (AI) can be stated as follows: we would like to generate samples of data drawn from a certain distribution $p(\phi)$, whose precise analytic form is not known but of which we already have (many) pre-existing samples. This is difficult, and one approach to the problem is to find a way to convert samples from a much simpler distribution (e.g. a Gaussian) to the samples that we want.


This is done by first constructing a probability flow from distributions of target data to a simpler distribution (usually a Gaussian distribution) can allow us to learn the inverse flow using a neural network, allowing us to go from pure Gaussian noise to generated data samples. In other words, we first learn how to destroy data by gradually turning it into noise. This is typically done through an Ornstein-Uhlenbeck process, which may be written in the form of an SDE:
\begin{equation}
d\phi = -\frac{\beta(t)}{2}\phi dt+\sqrt{\beta(t)}dW_t
\end{equation}
where $\phi(x,t)$ are the samples, and $t$ runs from $0$ at the data samples to some large positive value $T$ when the samples have been noised to approximately pure noise. 

The question is then: given that we have been able to destroy the samples, can we rebuild them starting from pure noise? We should then aim to find, given a sample, in which direction in probability space we want to go next in order to move towards a high-probability area of the data distribution. 

This is achieved through simple stochastic dynamics. Just as in normal thermodynamics we would want to enact the Langevin flow $dx = -\nabla U dt + dW_t$, where $U$ is the potential energy of the state, and could instead write this in terms of the Boltzmann distribution $p_0=\exp(-\beta U)$, very similar dynamics turn out to enact the reverse of the forward process in this case:
\begin{equation}
d\phi = -\frac{\beta(t)}{2}\phi dt - \beta(t) \nabla_x \log p_t(x) dt + \sqrt{\beta} dW_t
\end{equation}
Here $dW_t$ is noise and its coefficient $\sqrt{\beta}$ is chosen in such a way that the equilibrium distribution is exactly the Boltzmann distribution.

The important object to be able to carry out this reverse process computationally, and go from noise to data, is therefore the quantity $\pi(\phi,t) := \nabla_x \log p_t(x)$, which is the central learned object in diffusion models, and is called the \textit{score function}. In other words, we don't need to know the full normalized probability distribution we are aiming for at every point of the reverse process: we only need to know the gradient, or in which direction to flow next. 

In practice, the score function is learned using a neural network, using a method called score matching \citep{hyvarinen2005estimation} which is now commonly used for generative modeling \citep{song2019generative}. In particular, by discretising the forward (noising) process run on training data samples, the neural network can learn the score function for the reverse process. 

The specific form of the reverse process we use in this work, for field $\phi$, is the deterministic ODE form without noise:
\begin{equation}\label{eq:ODE_diff} 
\frac{\partial\phi(t)}{\partial t} = \frac{\partial_t \bar{\alpha}_t}{\bar{\alpha}_t} \left[\phi(t)+\pi\left(\phi(t),t\right)\right] 
\end{equation}
As explained in \cite{song2020score}, both this and the SDE form lead to the same data distribution at the end of the reverse process. In practice, the score function $\pi(\phi(t),t)$ is represented by a neural network which is trained on many samples from the data distribution. The reverse process starts from pure noise, so $\phi(0)$ is usually a grid of pointwise i.i.d. Gaussian variables; that is, $\phi(x,0) \sim \mathcal{N}(0,1)$ for all points $x$ within the sample.



\subsection{Microscopic derivation of creativity in local equivariant models}

Diffusion models such as described above work extremely well and are one of the workhorses of generative AI, generating seemingly creative examples of images, text, etc. As articulated clearly in \cite{kamb2025an}, it is curious to note that the training objective of such models should actually drive them to precisely {\it memorize} their training data: however when used in practice they instead generate creative outputs that seemingly remix their training data in interesting ways. Is it possible to obtain a universal theory for how this happens? 

A potential mechanism was given in \cite{kamb2025an}, where it was argued that in fact it is the inductive biases of the underlying neural networks which do not permit them to {\it precisely} fit the underlying objective and memorize their training data precisely. Rather they fail in an interesting and understandable manner. For example, when the network has a convolutional architecture, then the constraints of locality and translational equivariance imply that the score function which minimizes its training objective is given by an object called the {\it Equivariant Local Score Machine} or ELS: 
\begin{equation}
\pi[\phi](x) = -\frac{1}{1-\bar{\alpha}_t} \frac{\sum_{\varphi \in P_\Omega(\mathcal{D})}\left(\phi(x) - \sqrt{\bar{\alpha}_t} \varphi(0)\right)\mathcal{N}(\phi_{\Om_x} | \sqrt{\bar{\al}_t} \varphi, (1-\bar{\al}_t) I) }{\sum_{\varphi' \in P_{\Om}(\mathcal{D})} \mathcal{N}(\phi_{\Om_x} | \sqrt{\bar{\al}_t} \varphi' | (1- \bar{\al}_t) I)}  \label{ELSdef} 
\end{equation}
We briefly explain the objects arising in this formula: $\phi$ is the local field which is being evolved, and $\varphi$ runs over the set of all possible {\it patches} $P_{\Om}(\mathcal{D})$ taken from the training dataset $\mathcal{D}$. The size of each patch is a parameter $\Delta$ which must be specified. As explained in detail in \cite{kamb2025an}, this should be viewed as a mechanism by which each patch of the image is pulled towards an increasingly less noisy patch of the training data, resulting in a kind of mosaic of training data which manifests as a plausible and creative image. The mechanism above agrees closely on a sample-by-sample basis with the results of real learned models, and there is a sense in which it can be viewed as providing an {\it exact} solution to the problem of understanding training dynamics. 

We note however that while it is simple conceptually, in practice it is a rather complicated object: formulating the ELS requires an explicit sum over {\it all patches in the training data}. One can ask whether it is possible to understand the {\it general} dynamics of diffusion models in a simplified manner, ideally one in which the complexity of the training data appears only through a small number of parameters.  

\subsection{Effective field theory for the reverse process}

We now turn to the novel contribution of this work, i.e. the idea of using {\it effective field theory} (EFT) -- a concept originating in physics, but with wide applicability -- to achieve such a goal. Effective field theory is a formalization of the common-sense idea that nature can be understood at different {\it length scales}: e.g. one does not need to understand the microscopic structure of the proton (small scale) to study how water waves move (long scale). 

\subsubsection{Simple example of effective field theory}
At a technical level, this is usually accomplished by expanding various quantities order-by-order in derivatives. For the reader who is unfamiliar with this concept, we consider an extremely simple example. (The reader who is already familiar with EFT is encouraged to skip to the next subsection). Consider the following equation of motion for a particle moving in one dimension $x(t)$:
\be
\frac{d^2 x}{dt^2} + \tau \frac{d^3 x}{dt^{3}} = 0 \label{toy_eom} 
\ee
The first term is the normal movement of an idealized particle feeling no force. The second term is an example of a ``higher-derivative'' term that might appear in a more realistic model. Note that by dimensional analysis its coefficient $\tau$ must have dimensions of time. Let us study the most general solution to this third order differential equation, labeled by three integration constants $c_i$. 
\be
x(t) = c_1 + c_2 t + c_{3} e^{-\frac{t}{\tau}}
\ee
Now imagine that we are interested in the movement of the particle over {\it long} time-scales, i.e. with $t \gg \tau$. In this case the term in $c_{3}$ is exponentially small and can be neglected, and we simply obtain motion with a constant velocity. Note that this could perhaps have been anticipated {\it before} explicitly solving the equations of motion \eqref{toy_eom}: on general grounds, if we care about a regime where $t \gg \tau$, then we are saying that the time scale $\tau$ is very small compared to the time variation of the quantities of interest, and to good approximation we can set $\tau = 0$ from the outset, resulting in the same conclusion. Note that any further {\it higher} derivative terms (e.g. $\tau^{n-2}\frac{d^n}{dt^n} x(t)$) will have a coefficient proportional to higher powers of some quantity with dimensions of time $\tau$ and can also be neglected. This is a very powerful principle: essentially by dimensional analysis alone, the dynamics of almost any system {\it at suitably long time scales} is usually controlled by only a few parameters. The idea of {\it effective field theory} is to apply this general principle in a sophisticated manner, usually resulting in a controlled expansion that works for {\it long distances} or {\it late times}, often the regime of interest. The application of this idea to quantum field theory in particular is extremely well developed \citep{Weinberg:1978kz, Polchinski:1992ed}. See e.g. \citep{Burgess:2007pt,Kaplan:2005es} for reviews.  


\subsubsection{Effective theory of diffusion model} 
In this work we will apply the same approach to the buildup of correlations in a diffusion model with a convolutional backbone. It may not be obvious that the required principles apply in this setting. We will argue that the primary ideas required are {\it locality in space} and {\it translational equivariance}, are sufficient to allow us to use general principles of EFT to understand the reverse process. 

The basic idea behind our approach is the following. The dynamics of the reverse process is fully determined by the ODE \eqref{eq:ODE_diff}, in which the interesting dynamical input is the score function $\pi(\phi(t), t)$. In general, the score function $\pi(\phi(t),t)$ is an arbitrarily complicated functional of the current configuration $\phi(t)$, and it may seem impossible to understand how the model chooses to represent it. The insight of effective field theory is that this is often not the case: one can consider systematic expansions of the quantity of interest and then argue (as we did above in the toy example) that the interesting dynamics is determined by a few terms. 

To that end, let us consider a double expansion of $\pi(\phi(t),t)$ -- or, equivalently, the right hand side of the evolution equation \eqref{eq:ODE_diff} in a double expansion of $\phi(x,t)$ and {\it spatial derivatives} of $\phi(x,t)$. Now $\bar{\alpha}_t$ is a monotonic function of $t$, describing noise that is always being added as the forward process evolves. For example, in our implementation of the $\pm 1$ grid example, we use the standard cosine schedule from \cite{nichol2021improved}. This gives the following noise scheduling:
\begin{equation}
\bar{\alpha}_t = \prod_{t_i=0}^t \cos^2 \left(\frac{\pi}{2} \frac{\frac{t_i}{T}+s}{1+s}\right)
\end{equation}
Where $s$ is a small constant to ensure the absence of sharp zeroes or infinities, and $T$ is the final time in the forward process, with $t_i$ being the discretised time points along the forward process.

The result of this monotonicity is that we can always reparametrize time $t$ in terms of $\bar{\al}_t$, and we will do this from now on. 

Here are the first few terms of this expansion: 
\begin{align}
\frac{\partial \tilde{\phi}(x,\bar{\alpha}_t)}{\partial \bar{\alpha}_t} &= g_0(\bar{\alpha}_t) + g_2(\bar{\alpha}_t)\mathbf{v}_2\cdot \nabla \tilde{\phi}(x,t) + g_3 (\bar{\alpha}_t) \nabla^2 \tilde{\phi}(x,t)+  g_4\tilde{\phi}^2(x,t)+\nonumber \\ & + g_5 \tilde{\phi}^2 (x,t) \mathbf{v}_5 \cdot \nabla \tilde{\phi}(x,t) + g_6(\bar{\alpha}_t) \nabla \tilde{\phi}(x,t)\cdot \nabla \tilde{\phi}(x,t) + g_{7} \tilde{\phi}^3 + g_8 \tilde{\phi} \textbf{v}_8 \cdot \nabla \tilde{\phi} ... \label{eft_expans} 
\end{align}
where the $\mathbf{v}_i$ are vectors which do not depend on the field $\phi$ but (at this point) can have arbitrary dependence on time. 

We note a technical point: there is no term linear in $\phi$ on the right-hand-side of \eqref{eft_expans}. This is because such a term can be absorbed into a field redefinition. For example, if $a_1(\bar{\alpha}_t)$ were the coefficient of such a term, we could redefine $\tilde{\phi} = \exp\left[-\int^t dt' a_1(\bar{\alpha}_t)\right]\phi$ and absorb all resultant factors in the other terms into the definitions of their coefficients. Thus with no loss of generality we will assume that we have already done so and work with the rescaled field $\tilde{\phi}$ from now on\footnote{It is important that any observables of interest remain invariant under such a rescaling, which we will show to be true for the mutual information in our case}. 

Where do the two constraints of locality and translational equivariance enter in this expansion? Translational equivariance implies that the coefficient functions $g(\bar{\al}_t)$ cannot depend explicitly on space $x$. Locality implies that the score function at point $x$ depends more strongly on points which are close to the point $x$ than those that are far away, suggesting that a derivative expansion is possible in principle. However, for this expansion to also be {\it useful}, one has to meaningfully argue that that it can be truncated to a finite set of terms. 

To understand such a truncation, we first impose all of the symmetries of our underlying data distribution. We assume for this example that our underlying data distribution is rotationally equivariant, meaning that all of the vectors ${\bf v}_i = 0$. We also assume an underlying $\mathbb{Z}_2$ symmetry which acts as $\phi \to -\phi$, so that $g_{0}$, $g_{6}$ are zero. In that case we find the following much simpler expansion:
\begin{align}
\frac{\partial \tilde{\phi}(x,\bar{\alpha}_t)}{\partial \bar{\alpha}_t} &= g_3 (\bar{\alpha}_t) \nabla^2 \tilde{\phi}(x,t) + g_{7} \tilde{\phi}^3 + \cdots \label{eft_expans} 
\end{align}
We would like to consider dynamics near the pure noise endpoint at $\bar{\al}_t \to 0$. To that end we assume that all of the coefficient functions $g_{i}(\bar{\al}_t)$ have finite limits as $\bar{\al}_t \to 0$, i.e. $g_{i}(\bar{\al}_t \to 0) = c_i$, and furthermore we assume that in the regime of interest we can treat them as constant. 

We will now formalize the notion of studying ``long distances''. To do this we introduce a scaling limit, i.e. we consider zooming to longer distances (and later times) by introducing a parameter $\lam$. 
\be
\bar{\al}_t \to \lam^2 \bar{\al_t} \qquad x \to \lam x \qquad \lam \to \infty \label{lamscaling} 
\ee
One can now ask how the field $\tilde{\phi}$ is affected by this rescaling; as it turns out, when defining $\tilde{\phi}$ correctly one should assume the following scaling:
\be
\tilde{\phi} \to \lam^{-\frac{d}{2}} \tilde{\phi} \label{phidim} 
\ee
where $d$ is the number of spatial dimensions. We now briefly explain this formula, in two different ways. We first present a somewhat heuristic argument that is reminiscent of arguments in physics, and then explain an equivalent point of view arising from the central limit theorem. 

\ben \item What is the distribution of the original field $\phi(x,\bar{\al}_t = 0)$? At the start of the reverse process in diffusion this is a random variable drawn from a normal distribution of unit width, uncorrelated across the spatial sites. In other words, we have the following probability distribution:
\be
p(\phi(\bar{\al}_t = 0, x)) = \mathcal{N}^{-1} \prod_{i} \exp(-\frac{1}{2} \phi(x_{i})^2) \to \mathcal{N}^{-1} \exp\le(-\ha\int d^{d}x \phi(x)^2\ri)
\ee
where we have passed to a continuum notation in the second line. Now consider the rescaling $x \to \lam x$: in order for the distribution to remain invariant we should simultaneously rescale $\phi(x) \to \lam^{-\frac{d}{2}} \phi$. (This is precisely analogous to the usual procedure used to assign a scaling dimension to a quantum field in phyiscs, see e.g. the textbook treatment in \cite{Peskin:1995ev} ). 

\item This can also be understood from the central limit theorem. Let us consider an {\it averaged field} $\phi_{\ell}$ averaged over a cubic box of size $\ell$
\be
\phi_{\ell}(x) = \frac{1}{\ell^{d}} \sum_{y} \phi(y)
\ee
At $\bar{\al}_t = 0$, the sum runs over $\ell^{d}$ uncorrelated variables on each site, and thus has variance $\ell^{\frac{d}{2}}$. Thus the typical size of the averaged field is $\ell^{-\frac{d}{2}}$: rescaling $\ell \to \lam \ell$ now results in the same scaling \eqref{phidim}. Note that from this point of view we are studying not the original field $\phi$ but rather an averaged version $\phi_{\ell}$; this is the usual philosophy of EFT, and all such observables will agree with the original, as long as we study questions on scales that are significantly longer than $\ell$. 
\een 
The utility of establishing this scaling is that we can now use it to demonstrate which terms in the formal expansion \eqref{eft_expans} are important. Scaling all of the terms using \eqref{lamscaling} and \eqref{phidim}, we find:
\begin{align}
\lam^{-2-\frac{d}{2}} \frac{\partial \tilde{\phi}(x,\bar{\alpha}_t)}{\partial \bar{\alpha}_t} &= g_{3} \lam^{-2 -\frac{d}{2}} \nabla^2 \tilde{\phi}(x,t)+ g_{7} \lam^{-\frac{3d}{2}} \tilde{\phi}^3 + \cdots \label{eft_expans} 
\end{align}
We now take the limit $\lam \to \infty$. Importantly, any other terms which appear here with either more derivatives of higher powers of $\tilde{\phi}$ will come with a factor of $\lam$ raised to a {\it higher power} and can be neglected. In the case $d = 2$ the three terms shown above are the {\it only ones} which survive, and we thus find the following equation
\begin{align}
\frac{\partial \tilde{\phi}(x,\bar{\alpha}_t)}{\partial \bar{\alpha}_t} = g_{3} \nabla^2 \tilde{\phi}(x,t)+ g_{7} \tilde{\phi}^3 \label{eft_expans} 
\end{align}
which thus describes the growth of long-distance correlations shortly after the pure noise endpoint in a diffusion model in $d = 2$. 

This is as far as we can go using general physical principle of effective field theory. This is clearly an enormous simplification of the most general dynamics: the full microscopic dependence of the training dataset appears only through a few parameters, the $c_{i}$. As it turns out, we can go further still if we now use the microscopic description provided by the ELS in \eqref{ELSdef}. It is explained in detail in Appendix \ref{app:els_expansion} that in fact all nonlinearities of the form $\tilde{\phi}^{m}$ come multiplied with powers of $\bar{\alpha}_t$; to be precise, multiplied by a factor $\mathcal{O}(\bar{\al_t}^{\frac{m+1}{2}})$; thus the $\tilde{\phi}^3$ term can be neglected in the $\bar{\al}_t \to 0$ limit. This can be understood heuristically from the factors of $\sqrt{\bar{\al}_t} \varphi$ in \eqref{ELSdef}, which capture the fact that the mean of the distribution of the noised training data is drawn towards the origin at the noise endpoint. 
\footnote{We note that in the language of RG, a term in $c_{7} \tilde{\phi}^3$ would be {\it marginal}, which means that it would result in logarithmic corrections to all quantities that we compute. There is a well-developed framework to compute these corrections, but they would not have been visible in the finite resolution experiments that we perform in any case.} 

We thus find that the dynamics of the field close to the noise endpoint is given by the following extremely simple expression,
\be
\boxed{ \frac{\partial \tilde{\phi}(x,\bar{\alpha}_t)}{\partial \bar{\alpha}_t} = g_{3} \nabla^2 \tilde{\phi}(x,t)} \label{diffeqn} 
\ee
usually called the heat equation or diffusion equation. It is not a coincidence that it has appeared here: this equation is ubiquitous in describing physical phenomena for reasons ultimately arising from the effective field theory arguments given here. (Note the word ``diffusion'' plays multiple roles in this subject). 

\subsection{Mutual information growth}
We would now like to compute an observable using this simple model. 
In particular, we will study the growth of correlations at the early stages of the reverse process. One such measure of correlations is the mutual information between the field at two spatial points. We first define the connected two-point correlation function of the samples $\phi$ at time $t$ during the reverse process as follows:
\begin{equation}
G_t(x,y) := \langle \phi(x,t)\phi(y,t)\rangle - \langle\phi_t\rangle^2
\end{equation}
where the angled brackets denote an ensemble average (in this case, over multiple realisations of the initial noise). Note that we assume that $\vev{\phi(x,t)} = \vev{\phi_t}$ is constant in space by translational equivariance. 

We further assume that all quantities are Gaussian distributed; this is a self-consistent assumption as at $\al_{t} = 0$ the initial conditions are drawn from factorized Gaussians, and the linear evolution equation \eqref{diffeqn} will not introduce any higher-order moments. The mutual information between the lattice points $\phi(x)$ and $\phi(y)$ may then be written
\begin{equation}
I(x,y; t) = -\frac{1}{2} \log \left[1-\rho(x,y; t)^2\right]
\end{equation}
where $\rho(x,y;t)$ is the correlation coefficient of the distributions of the two pixels at time $t$:
\begin{equation}\label{eq:rho_definition}
\rho(x,y; t) := \frac{G_t(x,y)}{\langle \phi_t^2 \rangle - \langle \phi_t\rangle^2}
\end{equation}
Note, as claimed earlier, that the mutual information is therefore invariant under a field redefinition where $\phi$ is multiplied by a function of time, as in the EFT approach outlined above.

It is now possible to use the evolution equation \eqref{diffeqn} to explicitly solve for the growth of correlations as a function of diffusion time. The basic result is simple to state: at the beginning of the process each point is uncorrelated with the rest, and thus the mutual information is a delta function in space. As time goes on points become correlated and the mutual information itself diffuses outwards, becoming a Gaussian whose width grows with time. 

Such a Gaussian takes the following form
\begin{equation}\label{eq:gaussian_correlation_spread}
\rho(x,y;t) = \exp \left(-\frac{|x-y|^2}{4D\bar{\alpha}_t}\right)
\end{equation}
where $D$ is an effective diffusion constant that controls the rate of speed of the information. As we shall later see, this constant is exactly the coefficient of the EFT diffusion equation we study. We can see that as 'time' $\bar{\alpha}_t$ increases, the correlation, and therefore the mutual information, increases for the same separation $|x-y|$. Equivalently the correlation length $\xi$ defining mutual information growth grows diffusively as $\xi^2 \sim \bar{\al}_t$. 

The key result is that rather than depending on spatial separation and time independently, $\rho$, depends entirely on a self-similar, scale-invariant variable
\be
\rho(x,y;t) = \rho(|x-y|^2/\bar{\alpha}_t) \ . 
\ee 
A derivation of these results are provided in Appendix \ref{app:diffMI}. Therefore the EFT makes a clear prediction: at different times, if we plot against the self-similar variable $|x-y|^2/\bar{\alpha}_t$, we should obtain the same curve for $\rho$. In the later sections we will verify this prediction empirically in a few simple examples of trained diffusion models. 


\subsection{Snapping regime at end of reverse process}
Our discussion so far has been all near the beginning of the diffusion process. Here we briefly mention that an a different expansion can be performed near the {\it end} of the reverse process, where  $\bar{\alpha}_t \rightarrow 1^-$. Here we start from the form of the ELS in \cite{kamb2025an}, and expand in powers of $\epsilon_t := 1-\bar{\alpha}_t$. As the ELS involves a sum over patches $\varphi$ in the training data, we shall separate out:
\begin{flalign}
\frac{\partial \phi(x)}{\partial t} = -\frac{\partial_t \epsilon_t}{1-\epsilon_t} \frac{1}{\epsilon_t} \left((1-\epsilon_t) \phi(x) -\sqrt{1-\epsilon_t} \frac{\varphi^*(0)+\sum_{\varphi \in P_\Omega (\mathcal{D})}\varphi(0) e^{(\phi_{\Omega_x}-\sqrt{\bar{\alpha}_t}\varphi^*)^2-(\phi_{\Omega_x}-\sqrt{\bar{\alpha}_t}\varphi)^2/2\epsilon_t\textit{I}}}{1+\sum_{\varphi' \in P^y_\Omega (\mathcal{D})} e^{(\phi_{\Omega_x}-\sqrt{\bar{\alpha}_t}\varphi^*)^2-(\phi_{\Omega_y}-\sqrt{\bar{\alpha}_t}\varphi')^2/2\epsilon_t\textit{I}}}\right)
\end{flalign}
Thus, simplifying and dropping subleading terms in $\epsilon$ - which here are both $\mathcal{O}(\epsilon_t)$ and $\mathcal{O}(e^{-1/\epsilon_t})$ - we get a leading order evolution equation different to the diffusive regime, which now governs the end of the reverse process:
\begin{flalign}\label{eq:snapping_equation}
\frac{\partial \phi(x)}{\partial t} \approx -\frac{\partial_t \epsilon_t}{\epsilon_t} \left(\phi(x)-\varphi^*(0)\right)
\end{flalign}
This simple equation describes a simple exponential trajectory to the nearest training patch $\varphi^*$, and reflects the fact that as the reverse process nears its endpoint, individual patches decouple from each other and only proceed to the nearest training patch on an individual basis. This is the 'memorisation' regime in \cite{Biroli_2024}.

The limiting behaviour of the score function itself in this snapping regime is used in Theorem 4.1 of \cite{kamb2025an} to establish locally consistent points. We have here however further noted the explicit evolution equation that describes how the snapping happens dynamically, in the spirit of our EFT approach of describing dynamical behaviour with simple equations.

\section{Experiments}
We now show that the predictions of the effective field theory are borne out in two simple examples. 

\subsection{Two samples} 
We shall now demonstrate the diffusive regime quantitatively in a simple case with a full analytical description.

\begin{figure}
\centering
\includegraphics[width=\linewidth]{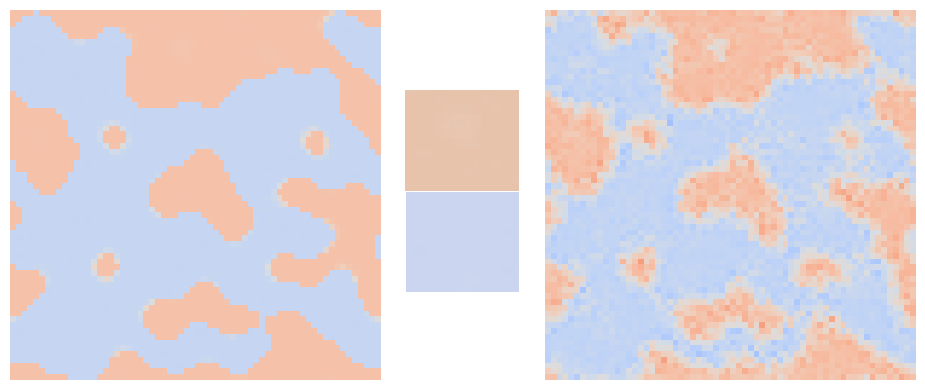}
\caption{Images generated at end of reverse process by analytical ELS machine (left) and trained ConvNet model (right). The training images are simply two homogenous grids of $\pm 1$ (centre).}
\label{fig:patches_in_els_and_model}
\end{figure}

Following \cite{kamb2025an}, we build a simple convolutional neural network for a diffusion model trained on only two samples: a grid where every pixel has the value $1$, and another grid where every pixel is $-1$. As explained in that work, due to the locality bias of the ConvNet, the diffusion model will generically produce patches with value $1$ and $-1$, of a characteristic size smaller than the size of the whole grid, as seen in Figure \ref{fig:patches_in_els_and_model}.
In this case, the ELS is easy to formulate, since it has only two possible patches in the training data: all $1$s, or all $-1$s. This leads to an explicit evolution equation of the form:
\begin{equation}\label{eq:ELS_ones_minus_ones_evolution_equation}
\frac{\partial \phi_t(x)}{\partial t} = \frac{\partial_t \bar{\alpha}_t}{\bar{\alpha}_t}\frac{1}{1-\bar{\alpha}_t}\left(-\bar{\alpha}_t \phi(x) +\sqrt{\bar{\alpha}_t}\tanh \left(\frac{\sqrt{\bar{\alpha}_t}\sum_{y \in \Omega_x}\phi(y)}{1-\bar{\alpha}_t}\right)\right)
\end{equation}

\begin{figure}
\centering
\begin{subfigure}{.47\textwidth}
  \includegraphics[width=\linewidth]{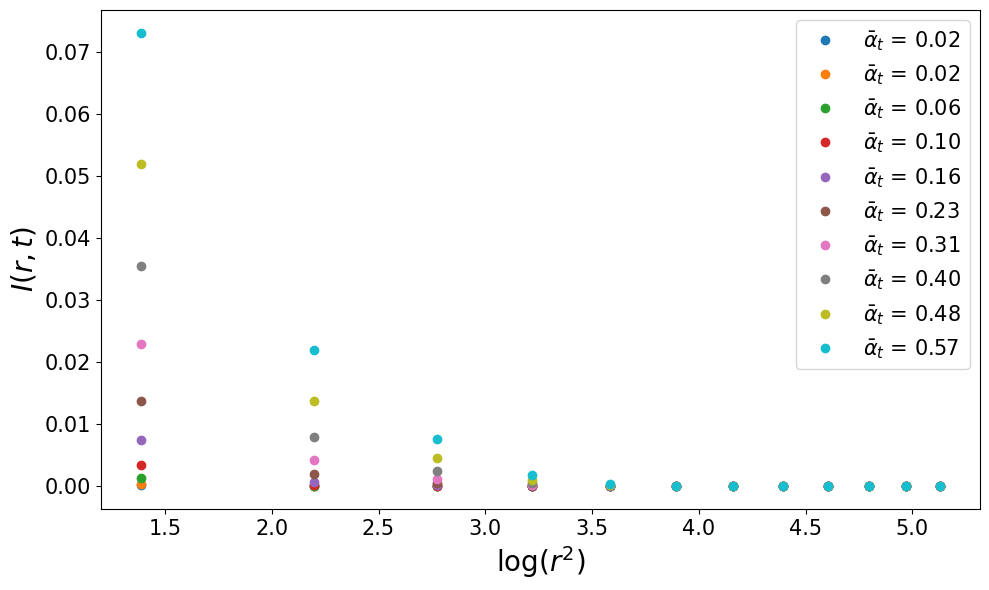}
  \caption{Mutual information curves against position at different times}
  \label{fig:sub1}
\end{subfigure}
\begin{subfigure}{.47\textwidth}
  \includegraphics[width=\linewidth]{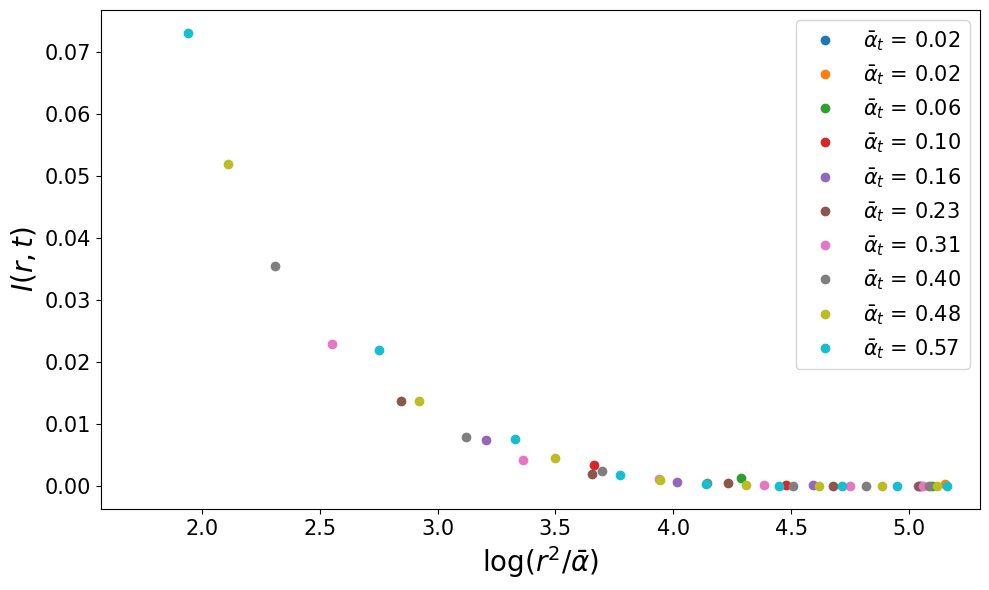}
  \caption{Mutual information curves against self-similar variable at different times collapse}
  \label{fig:sub2}
\end{subfigure}
\caption{Verification of diffusive behaviour of ELS machine with $5\times 5$ kernel size during reverse process.}
\label{fig:test}
\end{figure}

We plot the evolution of two-point correlations of the samples in the reverse process under the pure ELS (with a constant 5x5 locality kernel) against this self-similar variable in Figure \ref{fig:test}. This allows us to see that curves at different timesteps collapse, confirming our theoretical understanding of diffusive scaling.

\begin{figure}
\centering
\begin{subfigure}{.47\textwidth}
  \includegraphics[width=\linewidth]{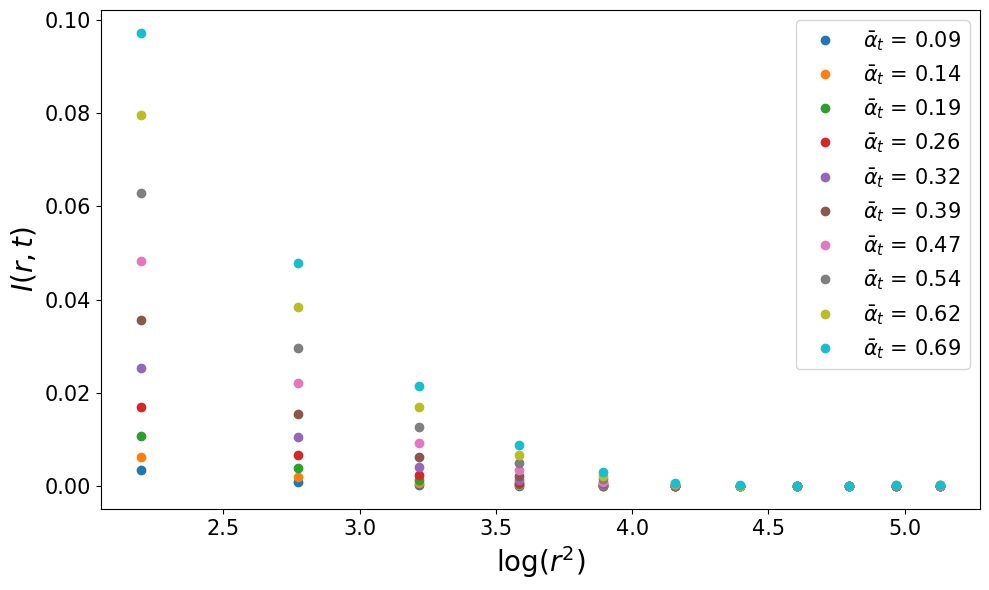}
  \caption{Mutual information curves against position at different times}
  \label{fig:sub1}
\end{subfigure}
\begin{subfigure}{.47\textwidth}
  \includegraphics[width=\linewidth]{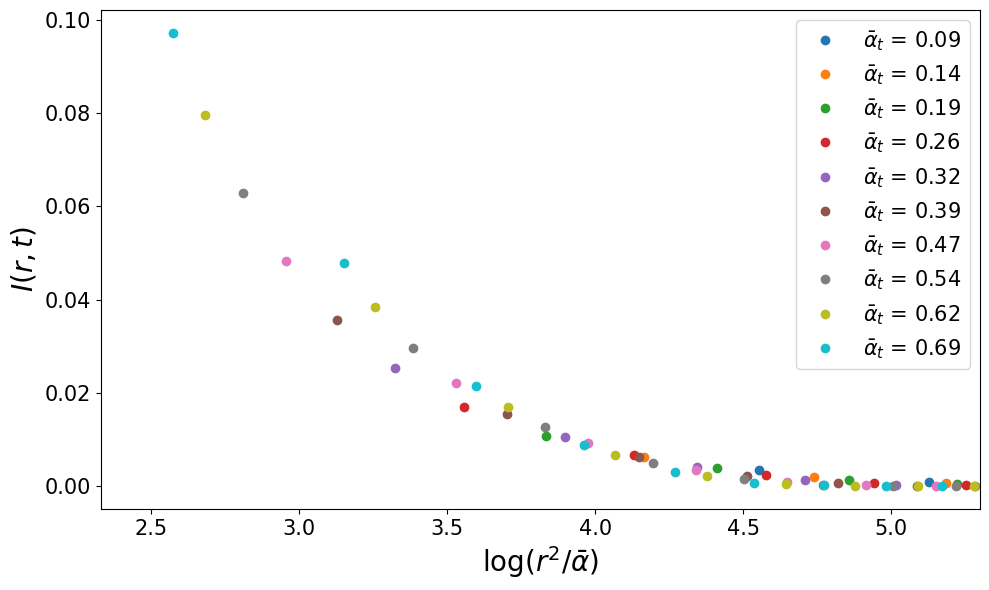}
  \caption{Mutual information curves against self-similar variable at different times collapse}
  \label{fig:sub2}
\end{subfigure}
\caption{Verification of diffusive behaviour of learned model during reverse process.}
\label{fig:actual_model_curve_collapse}
\end{figure}

This ELS machine can be compared directly to a simple diffusion model with a ConvNet backbone trained on an equally split batch of $1$ and $-1$ grids. Curve collapse extremely similar to the analytical ELS model may be observed in Figure \ref{fig:actual_model_curve_collapse}.
Note that we have excluded very small separations where $r$ is a few pixels, as at these scales, the results of the trained model diverge slightly from the pure ELS. This effect may be modelled by changing the size to illustrate our result.

\subsection{Transition between regimes in $\pm 1$ grids}
To understand the transition out of the diffusive regime, we can consider the action of the score function on a simple test grid $\phi_{\mathrm{test}}$, such as one which smoothly interpolates from $-1$ to $1$ along the $x_1$-axis (the horizontal axis, say) which runs along the interval $(-L/2,L/2)$:
\begin{equation}
\phi_{\mathrm{test}}(x_1,x_2)=\frac{2x_1}{L}
\end{equation}

Then, the score function becomes
\begin{equation}\label{eq:score_function_interp_input}
\pi[\phi_{\mathrm{test}}(x_1,x_2)]=-\frac{1}{1-\bar{\alpha}_t}\left(\left(\frac{2x_1}{L}\right) -\sqrt{\bar{\alpha}_t}\tanh \left(\frac{\sqrt{\bar{\alpha}_t}(2\Delta+1)^2\left(\frac{2x_1}{L}\right)}{1-\bar{\alpha}_t}\right)\right),
\end{equation}
where we note that the sum over the local patch has taken an elegant form here due to our simple test function.

If we now take the derivative of this quantity - the right hand side of the evolution equation, that determines how  at $x_1=L/2$, we obtain 
\begin{equation}\label{eq:ELS_ones_minus_ones_evolution_equation}
\frac{\partial \pi[\phi_{\mathrm{test}}(x_1,x_2)]}{\partial x_1}=\frac{2}{1-\bar{\alpha}_t}\left(-\frac{1}{L} +\frac{\bar{\alpha}_t(2\Delta+1)^2}{L(1-\bar{\alpha}_t)}\right)
\end{equation}
We understand that diffusion occurs when this is small and the evolution is flat, and stops occuring when a sharp spatial gradient forms at $x_1=0$, which means that the evolution is trying to snap even small positive or negative values to $+1$ or $-1$ respectively.

We need to define the transition by a factor $\Omega$. Let us mark it at the point when the absolute value of the spatial gradient of the score function at $x=0$, time $\bar{\alpha}_t$ is a factor $\Omega$ greater than that at the beginning of the reverse process $\bar{\alpha}_t=0$ (in this simple case, where the data is a straight line, that gradient is just $-2/L$). Here, we identify the end of the purely diffusive regime, as now nonlinearities are definitely important. 

Note that in this characterisation of the transition, there is no sharp, precise point where the behaviour changes from entirely diffusive to entirely snapping. Instead, there is always some amount of diffusive spreading of mutual information and some amount of snapping, and the best our method here allows us to do is to characterise this by the factor $\Omega$.

Applying this criterion in this case, and assuming that $\bar{\alpha}_t$ is small enough when the diffusive regime ends that we may approximate $1-\bar{\alpha}_t \approx 1$, we get a criterion:
\begin{equation}
\bar{\alpha}_t^{(crit)} \approx \frac{\Omega+1}{(2\Delta+1)^2} 
\end{equation}
Thus, the duration of the diffusive regime in this case is controlled by the effective ELS kernel size $\Delta$, or in more practical terms, solely by the architecture of the network. More generally, it is controlled by the structure of the training data and the network architecture.

However, one could conduct a similar procedure in the case of more complicated data by feeding similar test grids into the score functions. The general test grids would interpolate instead between local patches of the training data. The sharpness of the 'gradient' would be measured by dot-product overlap of the sample $\phi$ with these training patches. Note that there is no discontinuity in this transition, and the behaviour does not seem to correspond to usual physical notions of phase transitions.

Now, what is the correlation length developed at the end of the diffusive regime, and therefore, since no new spatial mutual information is developed in the rest of the reverse process, in the final samples? Recall the form of our EFT equation describing this diffusive regime:
\begin{equation}
\frac{\partial \tilde{\phi}(x,\bar{\alpha}_t)}{\partial \bar{\alpha}_t} = D \nabla^2 \tilde{\phi}(x,t)
\end{equation}
Here, $D$ is the diffusion constant that appears as the coefficient of the Laplacian term (which we earlier called, in our EFT series of terms, $g_3$). Recalling the Gaussian form of the correlation that this produces \eqref{eq:gaussian_correlation_spread}, and recalling that it evolves in space according to the variable $|x-y|^2/4D\bar{\alpha}_t$, we can correspondingly here write down the spatial correlation length $\xi$. This is just the characteristic decay lengthscale of this Gaussian - here, at time $\bar{\alpha}^{(crit)}_t$, it is
\begin{equation}
\xi \sim \sqrt{2D\bar{\alpha}^{(crit)}_t}\approx \frac{\sqrt{2D(\Omega+1)}}{2\Delta+1}
\end{equation}
However, in the simple $\pm 1$ grid example, we have a handle on how $D$ behaves as a function of the ELS patch size. Note that the sum over a locality kernel - the quantity that appears in the ELS in this case - is simply
\begin{equation}
\sum_{k,l=-\Delta}^\Delta \phi(x+k, y+l) 
\end{equation}
We wish to write this sum in terms of derivatives of $\phi$, which we can do by Taylor expanding. By symmetry, all first derivatives will vanish, as will the mixed derivative term $\partial_ x \partial_y \phi$.

We are interested, following our EFT approach, only in the first nontrivial term, which is the Laplacian term that induces diffusion. This will be the coefficient of $\partial_x^2 \phi$ (and also, by symmetry, of $\partial_y^2 \phi$). Noting that
\begin{equation}
\phi(x+k,y+l) = \phi(x,y)+...+\frac{k^2}{2} \partial_x^2 \phi(x,y)+\frac{l^2}{2} \partial_y^2 \phi(x,y)+...
\end{equation}

This yields the coefficient $D$ of $\partial_x^2 + \partial_y^2$:
\begin{equation}
D=\sum_{k,l=-\Delta}^\Delta \frac{k^2}{2} = \frac{1}{2} \left(\sum_{k=-\Delta}^\Delta k^2\right)\left(\sum_{l=-\Delta}^{\Delta} 1\right) = \frac{1}{2}\frac{\Delta(\Delta+1)(2\Delta+1)}{3} (2\Delta+1) \sim \Delta^4
\end{equation}
So, in the ELS,
\begin{equation}
\sum_{y\in \Omega_x} \phi(y) = \text{linear term in }\phi +\frac{1}{2}\frac{\Delta(\Delta+1)(2\Delta+1)}{3} (2\Delta+1) \nabla^2 \phi(x) +...
\end{equation}
This means that the achieved correlation length can be precisely written as 
\begin{equation}
\xi^{(crit)} \sim \frac{\sqrt{\frac{\Delta(\Delta+1)(2\Delta+1)}{3} (2\Delta+1)(\Omega+1)}}{2\Delta+1} \sim \Delta \sqrt{\Omega+1}
\end{equation}
Which means that the correlation length scales linearly with the ELS kernel size (as we would expect). If we do vary the ELS kernel size throughout the reverse process, we would expect the correlation length to be altered by a factor that is a function of the way the kernel size is varied.

Note that the 'correlation length' $\xi$ presented here is a statistical quantity derived from the averaged two-point function. In particular, at the end of the reverse process, it is different to the geometric correlation length, obtained by averaging the radii of clusters of $1$ or $-1$ in the produced samples. 

We would expect a similar result in the general case, but setting up a similar test function is analytically difficult since samples are in general not homogenous, and patches that interpolate between two dataset patches may be closer to a third patch already present in the dataset than either of the two being interpolated between, which requires careful handling that is not immediately clear.

We empirically test the theoretical computation of $D$ by plotting $\log \rho(r,t)$ against the self-similar variable $r^2/\bar{\alpha}_t$ for the real ConvNet data, since we expect $\rho$ to have a Gaussian shape. 
\begin{figure}
\centering
\includegraphics[width=0.7\linewidth]{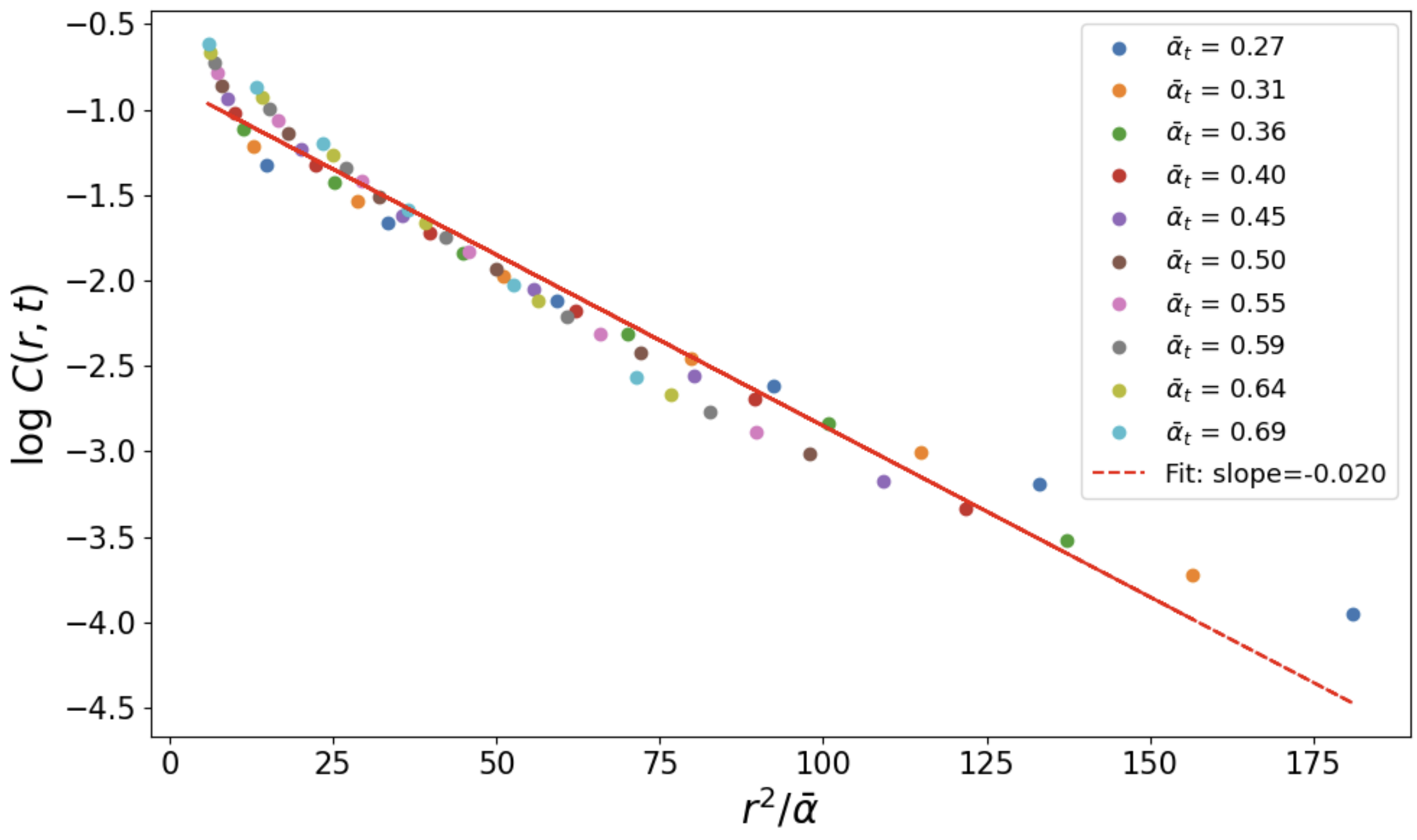}
\caption{Spatial dependence of $C(r,t)$; straight line with fixed slope $D$ illustrates Gaussian dependence.}
\label{fig:graph_check_eft_coeff}
\end{figure}
We see from Figure \ref{fig:graph_check_eft_coeff} that the gradient of the graph is $\approx 0.020$. According to theory, this gradient is $1/2D$, which gives $D \approx 25$. This is actually exactly consistent with a value of $\Delta = 2$. Naturally, we do not expect to get a perfect integer value of $\Delta$ since, as explained earlier, in reality the kernel size is shrinking throughout the reverse process. However, the correct order of magnitude provides a reassuring check of our theoretical calculations. 

Indeed, plotting these curves throughout the reverse process can also be a means of estimating $\Delta$, the kernel size, throughout the reverse process, giving a systematic inference method building on the reasoning in \cite{kamb2025an}. In the future, it would be interesting to investigate whether such a method carried out on test data, such as the $\pm 1$ grids here, can also help infer the variation of the kernel size throughout the reverse process for more complex real data. 

The overall picture we obtain therefore is as follows: close to the pure noise endpoint, the reverse process starts off by creating spatial mutual information on a scale which is the correlation length, which grows from 0, until it saturates at a fixed level set by the nonlinearity and the size of the locality kernel in the ELS machine.




\subsection{MNIST}
We now turn to a slightly more realistic example by training a diffusion model on the celebrated MNIST dataset. We first note that the justification of our theory discussion must be slightly modified, as MNIST does not have the symmetries we used earlier to obtain a simplified linear EFT equation.

\subsubsection{Relaxing symmetries}
The absence of the symmetries we earlier assumed means that in MNIST, and other realistic datasets, a number of other terms in the EFT will become significant. Here, we consider the first few linear terms of the EFT expansion
\begin{equation}
\frac{\partial \phi}{\partial \bar{\alpha}_t} = g_0(\bar{\alpha}_t) +g_1(\bar{\alpha}_t)\phi +g_2 \textbf{v}_2(\bar{\alpha}_t)\cdot\nabla \phi + g_3(\bar{\alpha}_t) \nabla^2 \phi
\end{equation}
Note that in this general case nonlinear terms in theory could appear in the expansion, such as $\phi^2, \phi \nabla \phi, \phi^3$. However, as we show in Appendix \ref{app:els_expansion}, for the ELS machine at least, each further power of $\phi^m$ brings with it a power of $\mathcal{O}(\bar{\alpha}_t^{(m+1)/2})$, which means that in the limit $\bar{\alpha}_t \rightarrow 0$ we can discard these terms in comparison with the linear terms we study here. We can redefine $\phi$ off the bat to make a number of the terms on the right hand side disappear. Consider the field redefinition:
\begin{equation}
\phi(x,\bar{\alpha}_t) = \exp\left(\int_{\bar{\alpha}_0}^{\bar{\alpha}_t} g_1(s) ds\right)  \tilde{\phi}\left(x-\int_{\bar{\alpha}_0}^{\bar{\alpha}_t} g_2 \textbf{v}_2(s) ds, \bar{\alpha}_t\right)
\end{equation}
Then the evolution equation for $\tilde{\phi}$ becomes
\begin{equation}
\frac{\partial \tilde{\phi}}{\partial \bar{\alpha}_t} = \exp\left(-\int_{\bar{\alpha}_0}^{\bar{\alpha}_t} g_1(s) ds\right) a(\bar{\alpha}_t) + g_3(\bar{\alpha}_t) \nabla^2 \tilde{\phi}
\end{equation}
From this simplified evolution equation it becomes clear that if we start with IID Gaussian random noise, that is,
\begin{equation}
\langle \phi(x,0)\rangle = 0, \langle \phi(x,0)\phi(y,0)\rangle = \delta(x-y)
\end{equation}
Then taking the expectation of the evolution equation after multiplying through by $\phi(y,t)$, and subtracting the appropriate one-point function terms, gives an evolution equation (as before) for the \textit{connected} two-point function of $\phi$:
\begin{equation}
\frac{\partial\langle \tilde{\phi}(x,t)\tilde{\phi}(y,t)\rangle_c}{\partial \bar{\alpha}_t} = 2g_3(\bar{\alpha}_t) \nabla^2\langle \tilde{\phi}(x,t)\tilde{\phi}(y,t)\rangle_c
\end{equation}
This has exactly the same solution as we derived before, if we introduce a new time co-ordinate $\tau(\bar{\alpha}_t) = \int_{\bar{\alpha}_0}^{\bar{\alpha}_t} ds \, g_3(s)$:
\begin{equation}
\langle \tilde{\phi}(x,\tau)\tilde{\phi}(y,\tau)\rangle_c = \frac{1}{(8\pi\tau)^{d/2}}\exp\left(-\frac{(x-y)^2}{8\tau}\right)
\end{equation}
But now, of course, the important new question in this case is what we get when we transform back to the connected two-point correlation function of the \textit{original} field $\phi$:
\begin{equation}
\langle \phi(x,\tau)\phi(y,\tau)\rangle_c = \frac{\exp\left(2\int_{\bar{\alpha}_0}^{\bar{\alpha}_t} g_1(s) ds\right)}{(8\pi\tau)^{d/2}}\exp\left(-\frac{(x-y)^2}{8\tau}\right)
\end{equation}
Where, rather nicely, because we only have a function of $x-y$, the shift of position in the definition of $\tilde{\phi}$ made no difference to the two-point function. And from here it is relatively straightforward to show that the mutual information is independent of the exponential prefactor. Therefore, once again, the $\nabla^2$ term - through $\tau$, which is defined from its coefficient - is the only one that actually controls the growth of correlations. 

\subsubsection{Experimental results for MNIST}
We now turn to the mutual information measurements. For simplicity we maintain circular padding, though this is not standard for image generation. Formulating the ELS with non-periodic boundary conditions, without circular padding, is also possible as detailed in \cite{kamb2025an}, but requires us to carefully distinguish between different types of edge and border local patches. 
\begin{figure}
\centering
\includegraphics[width=0.7\linewidth]{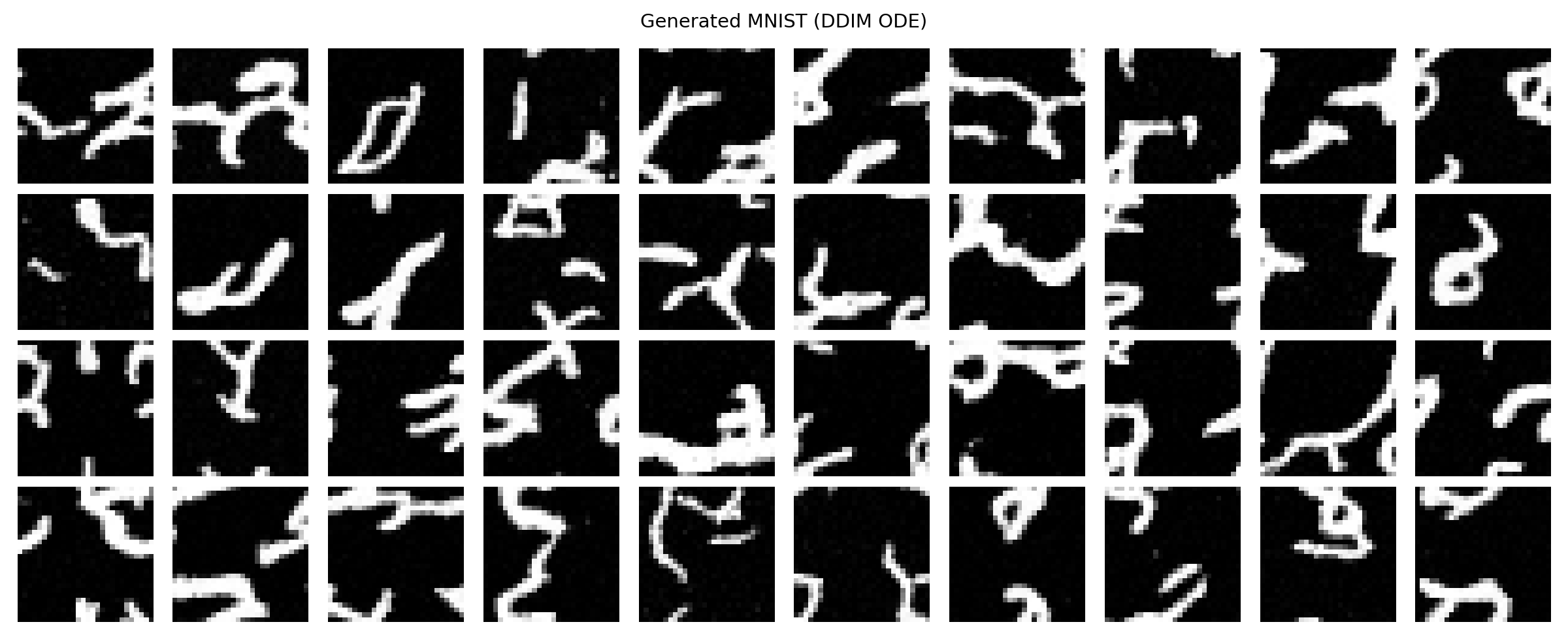}
\caption{Images generated by ConvNet with circular padding trained on MNIST dataset.}
\label{fig:mnist_generated_samples}
\end{figure}

\begin{figure}
\centering
\begin{subfigure}{.47\textwidth}
  \includegraphics[width=\linewidth]{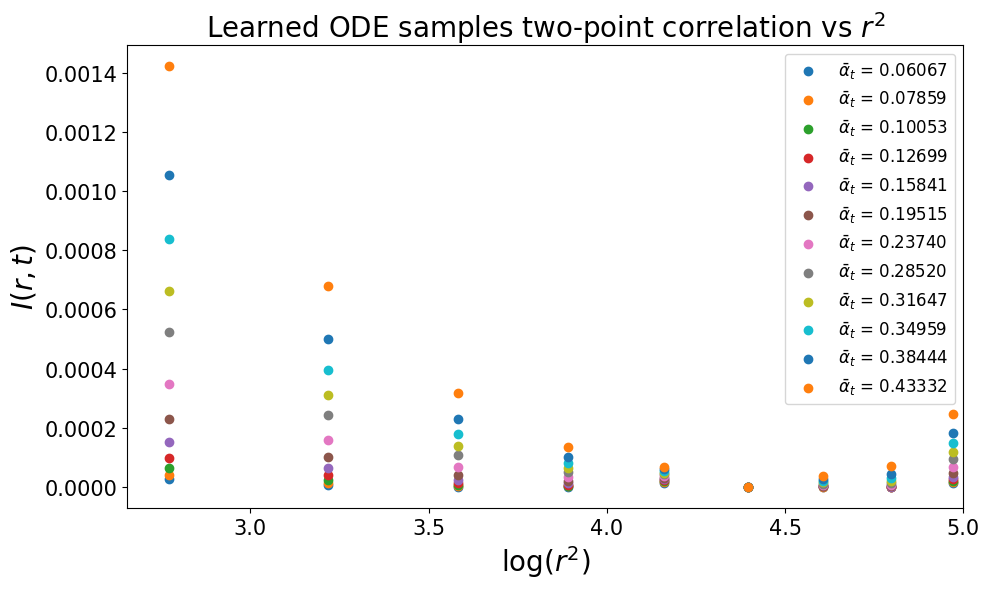}
  \caption{Mutual information curves against position at different times}
  \label{fig:sub1}
\end{subfigure}
\begin{subfigure}{.47\textwidth}
  \includegraphics[width=\linewidth]{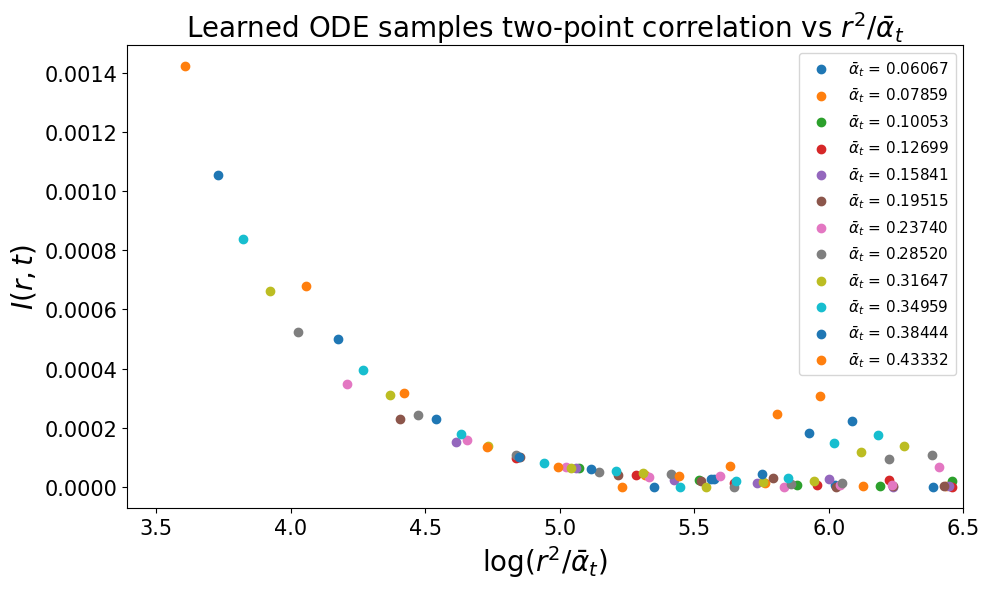}
  \caption{Mutual information curves against self-similar variable at different times collapse}
  \label{fig:sub2}
\end{subfigure}
\caption{Verification of diffusive behaviour of MNIST during reverse process. A 'tail' is visible where the correlation becomes negative due to the network learning a characteristic digit size.}
\label{fig:mnist_plots}
\end{figure}
The plots of mutual information are shown in Figure \ref{fig:mnist_plots}. In this figure, we again plot mutual information at scales beyond $r \sim$ a few pixels. We again see curve collapse, demonstrating agreement with the EFT at early times. 

Note that at large distances the collapse is not perfect. The large distance structure is sensitive to the choice of boundary conditions -- e.g. imposing an explicit boundary rather than using periodic boundary conditions changes the curves at large $r^2$ but also results in a reduced quality of data collapse. We believe this arises from an interaction of the diffusive modes with the boundaries. 


\subsection{Noise scheduling and noise in spatial correlations}

Our picture of the reverse process so far is split into the early diffusive regime, where spatial structure is built up, and the late 'snapping' regime, where local patches snap pointwise to the nearest patch in the training data.

We have seen that only the architecture of the network, and not the details of the precise noising schedule used, controls the spatial correlations built up in the final generated samples. However, the choice of noising does affect the 'fuzziness' of the spatial structure in the final samples. In the example of the $\pm 1$ grids, this manifests as noisiness in the colour of the patches.

This arises simply because samples during the reverse process are generated using a numerical integrator with evenly spaced timesteps. In our case, we have used a first-order Euler method everywhere. The approximate equation of motion for the snapping regime \eqref{eq:snapping_equation} is, pointwise, a first-order differential equation. This means that the error in the samples will be $\mathcal{O}(1/N)$ if there are $N$ steps in total during the snapping regime \footnote{Note that the step size $\delta t \propto 1/N$, so each step has a truncation error of $\delta t^2 \propto 1/N^2$. However, there are $N$ steps of each of these, leading to, in general, a final error of $\mathcal{O}(1/N)$}.

There is a different notion of 'fuzziness', which concerns the definition around the edges of regions of $\pm 1$. This is more difficult to treat analytically, requiring as it does a notion of 'region boundary' carried across various local patches. We leave the interesting possibility of such a theory to future work.

\section{Conclusion}
In this article, we have established the application of the formalism of EFT from physics to diffusion models with local convolutional architectures. We began with the idea that the learned score function in the reverse process, and therefore the evolution of the samples, can be expanded in a series of terms on general locality and symmetry principles, taking our cue from EFT expansions in physics.

The ELS machine of \cite{kamb2025an} provides a microscopic realisation of the score function learned by such a model. We showed that our EFT approach can vastly simplify the complicated analytic structure of the ELS machine (at least, near the limiting endpoints of the reverse process) into equations involving a few straightforward terms, and allow us to study the growth of spatial mutual information in samples over the reverse process using these simple EFT equations.

The EFT allowed us to argue for the restriction of the evolution equation to a few simple terms, but these also included nonlinear terms, which are difficult to keep track of analytically. Happily, the structure of the ELS showed that these nonlinear terms are unimportant in the regime of interest (near the pure noise endpoint). We also calculated these equations explicitly from the ELS for a toy model of $\pm 1$ grids.

Our work has characterised the existence of regimes within the reverse process, together with a quantitative description of their dynamics. In particular, in experiments on both the $\pm 1$ grids and on MNIST, we have directly shown - in accordance with our theoretical predictions - the existence of a diffusive regime where information spreads in space according to the self-similar variable $x^2/\bar{\alpha}_t$, so that the noise schedule plays the role of time in the diffusion equation.

It is interesting to ask whether an effective field theory approach could be helpful for other network architectures, e.g. transformers. In particular, it has recently been argued that a notion of locality is inherited from the {\it data}, independent of the underlying network architecture \cite{lukoianov2025locality}; we leave this interesting topic for further research. 


\subsubsection*{Acknowledgments}
We are grateful to L. Ambrogioni for discussions and comments on the draft. This work was supported by a grant from the
Simons Foundation (PD-Pivot Fellow-00004147, NI). NI and NN are supported by the STFC under grant number ST/T000708/1. Claude Code was used to implement experiments, collect data and visualise the results.

\bibliography{all}
\bibliographystyle{iclr2026_conference}

\appendix
\section{Appendix}
\subsection{Long-distance description of equivariant local score machine}\label{app:els_expansion}
Here we study the equivariant local score machine discussed in \cite{kamb2025an} and perform a systematic expansion in derivatives, showing that it admits a weakly-coupled fixed point near the pure-noise endpoint. 

The starting point is Eq (9) of \cite{kamb2025an}, which provides the following expression for the score $M_t[\phi](x)$
\be
M_{t}[\phi](x) = \sum_{\varphi \in P_{\Om}(\mathcal{D})} \frac{\le(\sqrt{\bar{\al}_t} \varphi(x) - \phi(x)\ri)}{1-\bar{\al}_t} W_{t}(\varphi | \phi, x) \label{eq:els_def} 
\ee
where the weight $W_{t}(\varphi | \phi, x)$ is defined to be:
\be\label{eq:weightdef}
W_{t}(\varphi | \phi, x) = \frac{\mathcal{N}(\phi_{\Om_x} | \sqrt{\bar{\al}_t} \varphi, (1-\bar{\al}_t) I)}{\sum_{\varphi' \in P_{\Om}(\mathcal{D})} \mathcal{N}(\phi_{\Om_x} | \sqrt{\bar{\al}_t} \varphi' | (1- \bar{\al}_t) I)}  
\ee
and where the normal distribution can be written explicitly as
\be
\mathcal{N}(\phi_{\Om_x} | \sqrt{\bar{\al}_t} \varphi, (1-\bar{\al}_t)I) = n \exp\le[-\frac{1}{2}(1-\bar{\al}_t)^{-1}\sum_{y \in {\Om_x}}  \le(\phi(y) - \sqrt{\bar{\al}_t} \varphi(y)\ri)^2\ri]
\ee
(where the normalization constant $n$ cancels between numerator and denominator, and thus we do not calculate it). Note that the sum over $y$ is over the patch $\Om_{x}$ centered about $x$. We have used a notation slightly different from \cite{kamb2025an} in that we write $\varphi(x)$ rather than $\varphi(0)$ to denote the value of the datapoint in question at the center of the patch $\Om_{x}$. 

We now adopt a continuum notation, i.e.
\be
\sum_{y \in \Om_{x}} \to \int_{\Om_{x}} dy
\ee
and systematically expand the right-hand side of Eq. \eqref{els_def} in derivatives of $\phi$ about the evaluation point $x$. This is possible because it depends only on {\it patches} $\Om_{x}$, and is thus a short-range kernel; such a short range kernel always admits an analytic expansion in momentum space and can thus be expanded in derivatives. This expansion is generally only {\it useful} when the derivatives are themselves small and thus the expansion can be truncated at some order. 

Our main point is that this expansion can always be systematically written in terms of an expansion in powers of $\phi$ and derivatives of $\phi$. The first few terms in such an expansion take the form:
\be
M_{t}[\phi](x) = a_{0,0}(\bar{\al}_t) + a_{1,0}(\bar{\al}_t) \phi(x) + a_{1,1}^{\mu}(\bar{\al}_t)\p_{\mu}\phi(x) + a_{1,2}^{\mu\nu}(\bar{\al}_t) \p_{\mu} \p_{\nu}\phi(x) + \cdots
\label{coeffdef} \ee
where a term in $a_{m,n}$ multiplies $m$ powers of $\phi$ and $n$ derivatives of $\phi$. We will explicitly work out only the first few terms as the expansion rapidly becomes unwieldy. 

To proceed we write 
\be
\phi(y) = \phi(x) + (y-x)^{\mu} \p_{\mu} \phi(x) + \ha (y-x)^{\mu}(y-x)^{\nu} \p_{\mu}\p_{\nu} \phi(x) + \cdots
\ee
and insert this into the integral in the exponent to find
\be
\frac{1}{2} \int_{\Om_{x}} dy \le(\phi(y) - \sqrt{\bar{\al}_t} \varphi(y)\ri)^2 = \frac{1}{2} \int_{\Om_{x}} dy \le(\phi(x) - \sqrt{\bar{\al}_t} \varphi(y) + (y-x)^{\mu} \p_{\mu} \phi(x) + \ha (y-x)^{\mu}(y-x)^{\nu} \p_{\mu}\p_{\nu} \phi(x) + \cdots\ri)^2
\ee
We can now systematically expand in derivatives. We first expand the normal distribution in an expansion as
\be
\mathcal{N}(\phi_{\Om_x} | \sqrt{\bar{\al}_t} \varphi, (1-\bar{\al}_t)I) = b_{0,0} + b_{1,0} \phi(x) + b_{1,1}^{\mu}\p_{\mu}\phi(x) + b_{1,2}^{\mu\nu}\p_{\mu} \p_{\nu}\phi(x) + \cdots
\ee
where the expansion coefficients $b_{m,n}[\bar{\al}_t, \varphi]$ implicitly depend on $\bar{\al}_t$ and the data point in question $\varphi$. All of the terms without derivatives can be written as:
\be
\sum_{m=0}^{\infty} b_{m,0} \phi(x)^{m} = n \exp\le[-\frac{1}{2}(1-\bar{\al}_t)^{-1} \int dy \le[\phi(x) - \sqrt{\bar{\al}_t} \varphi(y)\ri]^2\ri]
\ee
we may thus read off
\be
b_{0,0} = n \exp\le[-\frac{1}{2}(1-\bar{\al}_t)^{-1} \int dy\le[\sqrt{\bar{\al}_t} \varphi(y)\ri]^2\ri]
\ee
etc. and similarly for all $b_{m,0}$. We write down the first few derivative terms that are linear in $\phi$. Explicitly, we have
\begin{align}
b_{1,1}^{\mu} & = b_{0,0}(1-\bar{\al}_t)^{-1} \int_{\Om_{x}} dy (y-x)^{\mu}\sqrt{\bar{\al}_t} \varphi(y) \\
b_{1,2}^{\mu\nu} & = \frac{1}{2} b_{0,0}(1-\bar{\al}_t)^{-1} \int_{\Om_{x}} dy (y-x)^{\mu}(y-x)^{\nu} \sqrt{\bar{\al}_t} \varphi(y)
\end{align}
These expansion coefficients are somewhat analogous to familiar expressions ``moments of inertia'', where the data variables $\varphi$ play the role of a density.  

We now assemble the pieces. For notational convenience denote
\be
Z[\bar{\al}_t] \equiv \sum_{\varphi' \in P_{\Om}(\mathcal{D})} b_{0,0}[\bar{\al}_t,\varphi']
\ee
The denominator of \eqref{eq:weightdef} then becomes
\be
\sum_{\varphi' \in P_{\Om}(\mathcal{D})} \mathcal{N}(\phi_{\Om_x} | \sqrt{\bar{\al}_t} \varphi' , (1- \bar{\al}_t) I) = Z\le(1 + Z^{-1} b_{1,0} \phi + Z^{-1} \sum_{\varphi'} b_{1,1}^{\mu}[\varphi'] \p_{\mu}\phi(x) + Z^{-1} \sum_{\varphi'} b_{1,2}^{\mu\nu}[\varphi']\p_{\mu}\p_{\nu}\phi(x)+ \cdots\ri)
\ee
and if we introduce the following notation for a function $a[\varphi]$ of data points $\varphi$:
\be
[a]_{c} \equiv a(\varphi) - Z^{-1} \sum_{\varphi' \in P_{\Om}(\mathcal{D})}a(\varphi')
\ee
then the weight \eqref{eq:weightdef} can be written
\be
W_{t}(\varphi | \phi, x) = \frac{1}{Z} \le( b_{00} + [b_{1,0}(\varphi)]_{c} \phi + [b^{\mu}_{1,1}(\varphi)]_{c} \p_{\mu}\phi + \cdots\ri)
\ee
Inserting this into \eqref{els_def} and expanding out the terms, we can finally obtain explicit expressions for the first few coefficients in \eqref{coeffdef}:
\begin{align}
a_{00} & = \frac{1}{Z}\frac{1}{1-\bar{\al}_t} \sum_{\varphi} \sqrt{\bar{\al}_t} \varphi b_{00}(\varphi) \\
a_{1,0} & = \frac{1}{Z}\frac{1}{1-\bar{\al}_t} \sum_{\varphi} \le(\sqrt{\bar{\al}_t} \varphi [b_{1,0}]_{c}(\varphi) - b_{0,0}\ri) \\
a_{1,1}^{\mu} & = \frac{1}{Z}\frac{1}{1-\bar{\al}_t} \sum_{\varphi} \le(\sqrt{\bar{\al}_t}\varphi [b^{\mu}_{1,1}]_{c}(\varphi)\ri) \\
a_{1,2}^{\mu\nu} & = \frac{1}{Z}\frac{1}{1-\bar{\al}_t} \sum_{\varphi} \le(\sqrt{\bar{\al}_t}\varphi [b^{\mu\nu}_{1,2}]_{c}(\varphi)\ri)
\end{align}
These expressions are lengthy and not really very illuminating. There are two important points here: the first is really just to state that given a microscopic description (in this case the equivariant local score machine), explicit formulas can be found for the terms in the effective field theory expansion (in this case, writable as sums over the dataset). Furthermore, often symmetries are useful in constraining these terms. For example, imagine that the underlying dataset has a ${Z}_2$ symmetry which takes $\phi \to -\phi$: if the underlying neural networks are equivariant under this symmetry (\cite{cohen2016group}), then we can see that $a_{00} = 0$. Similarly, if the dataset has rotational symmetry then we can set $a^{\mu}_{1,1} = 0$, and $a^{\mu}_{1,2} = \delta^{\mu\nu}$ as the only rotationally invariant two-tensor.

The second important point considers the scaling of terms in this expansion. Note that in the formula for the EFT \eqref{eq:els_def}, the weight contains a numerator and denominator both of which contain an $\exp(-\phi_{\Omega_x}^2)$ term that can be cancelled. In other words, we can also write the weight as
\begin{equation}
W_t(\varphi | \phi, x) = \frac{\exp\le[-\frac{1}{2}(1-\bar{\al}_t)^{-1} \int dy \le(-2\phi(x) \sqrt{\bar{\al}_t} \varphi(y)+\bar{\alpha}_t \varphi(y)^2\ri)\ri]}{\sum_{\varphi' \in P_{\Om}(\mathcal{D})} \exp\le[-\frac{1}{2}(1-\bar{\al}_t)^{-1} \int dy \le(-2\phi(x) \sqrt{\bar{\al}_t} \varphi'(y)+\bar{\alpha}_t \varphi'(y)^2\ri)\ri]}
\end{equation}
From this rearrangement, it is clear to see that every time $\phi(x)$ appears, it is multiplied by a factor of $\sqrt{\bar{\alpha}_t}$, so that $W_t(\varphi | \phi, x)$ is really a function $W_t(\varphi | \sqrt{\bar{\alpha}_t}\phi, x)$. 

Now, before we put the weight back into the score, note that the evolution equation itself \eqref{eq:ODE_diff}, with the ELS score function, is
\begin{align}
\frac{\partial\phi(t)}{\partial \bar{\alpha}_t} &= \frac{1}{\bar{\alpha}_t} \left[\phi(t)+\pi\left(\phi(t),t\right)\right] \nonumber \\ &= \frac{1}{\bar{\alpha}_t}\sum_{\varphi \in P_{\Om}(\mathcal{D})} \frac{\le(\sqrt{\bar{\al}_t} \varphi(x) - \bar{\alpha}_t\phi(x)\ri)}{1-\bar{\al}_t} W_{t}(\varphi | \phi, x) \label{els_def} 
\end{align}

Putting the weight back into the full equation \eqref{eq:els_def} for the score function, therefore, we see that
\begin{equation}
a_{m,0} = \mathcal{O}(\bar{\alpha}_t^{(m+1)/2})
\end{equation}
simply because the weight is a function of $\sqrt{\bar{\alpha}_t}\phi$: every new term involving a new power of $\phi$ in the expansion comes with at least a power of $\bar{\alpha}_t^{1/2}$.

This means that nonlinear terms such as $\phi^2, \phi^3, \phi \nabla \phi$ in our EFT expansion become strictly subleading (in powers of $\bar{\alpha}_t$, in the limit $\bar{\alpha}_t \rightarrow 0$) to linear terms. This is why they have no visible impact in the experimental data we see.

We verify this approach for the $\pm 1$ grid example to show that EFT coefficients may be meaningfully extracted using these methods. To denote the $+1$ grid, we shall write $\varphi^+(x) = 1$ for all $x$, and similarly $\varphi^-(x) = -1$ everywhere. First, we calculate
\begin{equation}
b_{0,0}[\bar{\alpha}_t, \varphi^+] = b_{0,0}[\bar{\alpha}_t, \varphi^-] = n \exp\le[-\frac{1}{2}\bar{\al}_t(1-\bar{\al}_t)^{-1} \int_{\Omega_x} dy\ri]
\end{equation}
We are going to keep the integral explicit for now, since it is technically an approximation of a discrete sum to which we shall find it useful to revert later.

Then, the quantity $Z[\bar{\alpha}_t]$ in this case is:
\begin{equation}
Z[\bar{\alpha}_t] = \sum_{\varphi' \in P_{\Om}(\mathcal{D})} b_{0,0}[\bar{\al}_t,\varphi'] = 2n \exp\le[-\frac{1}{2}\bar{\al}_t(1-\bar{\al}_t)^{-1} \int_{\Omega_x} dy\ri]
\end{equation}
Next, we need to work out
\begin{align}
b_{1,2}^{\mu\nu}[\varphi^+] & = \frac{1}{2}b_{0,0}(1-\bar{\al}_t)^{-1}\sqrt{\bar{\al}_t} \int_{\Om_{x}} dy (y-x)^{\mu}(y-x)^{\nu} 
\end{align}
And similarly, we'll have 
\begin{align}
b_{1,2}^{\mu\nu}[\varphi^-] & = -\frac{1}{2}b_{0,0}(1-\bar{\al}_t)^{-1}\sqrt{\bar{\al}_t} \int_{\Om_{x}} dy (y-x)^{\mu}(y-x)^{\nu} 
\end{align}
Now to proceed to working out $a^{\mu\nu}_{1,2}$, which is the quantity we want, note that 
\begin{equation}
Z^{-1} \sum_{\varphi' \in P_{\Om}(\mathcal{D})}b_{1,2}^{\mu\nu}(\varphi') = b_{1,2}^{\mu\nu}[\varphi^+] + b_{1,2}^{\mu\nu}[\varphi^-] =0.
\end{equation}
So 
\begin{align}
[b_{1,2}]_c^{\mu\nu}(\varphi^+) & = b_{1,2}^{\mu\nu}(\varphi^+)
\end{align}
Then, we can finally write
\begin{align}
a_{1,2}^{\mu\nu} & = \frac{1}{Z}\frac{1}{1-\bar{\al}_t} \sum_{\varphi} \le(\sqrt{\bar{\al}_t}\varphi [b^{\mu\nu}_{1,2}]_{c}(\varphi)\ri) \nonumber \\ &= \frac{1}{Z}\frac{1}{1-\bar{\al}_t} b_{0,0}(1-\bar{\al}_t)^{-1}\bar{\al}_t \int_{\Om_{x}} dy (y-x)^{\mu}(y-x)^{\nu} \nonumber \\ &=  \frac{\bar{\al}_t}{2(1-\bar{\al}_t)^2} \int_{\Om_{x}} dy (y-x)^{\mu}(y-x)^{\nu}
\end{align}
If we now consider $\Omega_x$ as a $3x3$ cell, and evaluate this integral at face value, we get
\begin{align}
a_{1,2}^{\mu\nu} & = \frac{2\bar{\al}_t\delta^{\mu\nu}}{(1-\bar{\al}_t)^2} 
\end{align}
However note that this is only an approximation to the true discrete sum, which works out instead over a $3\times3$ grid to
\begin{align}\label{eq:nabil_method_eft_coeff}
a_{1,2}^{\mu\nu} & = \frac{3\bar{\al}_t\delta^{\mu\nu}}{(1-\bar{\al}_t)^2} 
\end{align}
As a check of the explicit EFT formalism above, we next present a different derivation of the same EFT coefficient, from first principles only from our analytic description of the $\pm 1$ grid case, which provides a robust check on its validity. We shall see that it produces exactly the same coefficient as our discrete calculation above.
\subsection{Checking against explicit ELS expansion in two sample case}
Here instead, we'd have 
\begin{flalign}
\pi(\phi,t)(x) = -\frac{1}{(1-\bar{\alpha}_t)} \left[\phi(x) - \sqrt{\bar{\alpha}_t} \tanh \left(\frac{\sqrt{\bar{\alpha}_t}\sum_{y \in \Omega_x}\phi(y)}{1-\bar{\alpha}_t}\right)\right]
\end{flalign}
So the differential equation for the reverse process is
\begin{equation}
\frac{d\phi(x)}{dt} = \frac{\partial_t \bar{\alpha}_t}{\bar{\alpha}_t}\frac{1}{1-\bar{\alpha}_t}\left(-\bar{\alpha}_t \phi(x) +\sqrt{\bar{\alpha}_t}\tanh \left(\frac{\sqrt{\bar{\alpha}_t}\sum_{y \in \Omega_x}\phi(y)}{1-\bar{\alpha}_t}\right)\right)
\end{equation}


Let's assume to start with, as in the previous section for the sake of comparison, that the neural network's 'locality' patches are 3x3 squares around each pixel. Generalisations to higher dimension locality patches will just involve higher order derivatives.

Note firstly that 
\begin{eqnarray}
\phi_{x+\Delta,y} + \phi_{x-\Delta,y} \approx \nabla_x^2 \phi_{x,y} + 2 \phi_{x,y}
\end{eqnarray}
In the continuum limit. A similar expression can be written down for $y$, and to take account of the diagonal squares in the patch, note
\begin{eqnarray}
\phi_{x+\Delta,y+\Delta} + \phi_{x-\Delta,y-\Delta} \approx (\boldsymbol{n}_1 \cdot \nabla)^2 \phi_{x,y} + 2 \phi_{x,y} \\ \phi_{x+\Delta,y-\Delta} + \phi_{x-\Delta,y+\Delta} \approx (\boldsymbol{n}_2 \cdot \nabla)^2 \phi_{x,y} + 2 \phi_{x,y} 
\end{eqnarray}
Where $\boldsymbol{n}_1:=(1,1), \boldsymbol{n}_1:=(1,-1)$.
Summing all these together, therefore, we'll get just (where $\Omega^3$ indicates that we've specialised to the case of a 3-by-3 locality kernel)
\begin{equation}
\sum_{y \in \Omega^3_x}\phi(y) \approx 3 \nabla^2 \phi(x) + 9 \phi(x)
\end{equation}
(Note that we did not need to assume commutation of derivatives here. A quick check with $\boldsymbol{n}_1\cdot \nabla, \boldsymbol{n}_1\cdot \nabla$ in the above suffices to prove this. )

Therefore we can now expand this term in the Taylor series of the $\tanh$. The evolution equation becomes
\begin{flalign}
\frac{d\phi(x)}{dt} \approx \frac{\partial_t \bar{\alpha}_t}{\bar{\alpha}_t}\frac{1}{1-\bar{\alpha}_t}\left(-\bar{\alpha}_t \phi(x) +\sqrt{\bar{\alpha}_t}\tanh \left(\frac{9\sqrt{\bar{\alpha}_t}\phi(x)}{1-\bar{\alpha}_t}\right) + \frac{3\bar{\alpha}_t \nabla^2 \phi(x)}{1-\bar{\alpha}_t} \text{sech}^2\left(\frac{9\sqrt{\bar{\alpha}_t}\phi(x)}{1-\bar{\alpha}_t}\right) + ...\right)
\end{flalign}

Now when $\bar{\alpha}_t \approx 0$ near the beginning of the reverse process,
\begin{flalign}
\frac{d\phi(x)}{dt} \approx \frac{\partial_t \bar{\alpha}_t}{\bar{\alpha}_t}\frac{1}{1-\bar{\alpha}_t}\left(-\bar{\alpha}_t \phi(x) +\bar{\alpha}_t\frac{9\phi(x)}{1-\bar{\alpha}_t} + \frac{3\bar{\alpha}_t \nabla^2 \phi(x)}{1-\bar{\alpha}_t}\right)
\end{flalign}
Note that this is a consistent expansion because all higher terms are $\mathcal{O}(\bar{\alpha}_t^{3/2})$, which means they will disappear compared to these. 


After rewriting, in the limit as $\bar{\alpha}_t \rightarrow 0$, we can formulate the equation much more simply:
\begin{flalign}
\frac{\partial\tilde{\phi}(x,\bar{\alpha}_t)}{\partial \bar{\alpha}_t} \approx 3 \nabla^2 \tilde{\phi}(x, \bar{\alpha}_t)
\end{flalign}
Where we've reparametrised time $t \rightarrow \bar{\alpha}_t$, and where
\begin{equation}
\tilde{\phi} = e^{-8 \bar{\alpha}_t} \phi
\end{equation}
Note this time prefactor in front doesn't mean much: it doesn't affect the positional part of the solution, which determines the behaviour of spatial correlations.

The important number is the coefficient $3\bar{\alpha}_t$ in front of the diffusion term, which is in precise agreement with the answer computed using the previous method \eqref{eq:nabil_method_eft_coeff}.

Indeed, we can immediately read off the characteristic length scale (normally, $\xi \sim \sqrt{2Dt}$ for a diffusion equation with diffusion coefficient $D$ - see the following appendix for a derivation):
\begin{flalign}
\xi(t) \sim \sqrt{6 \bar{\alpha}_t}
\end{flalign}

\section{Diffusion equation evolution and scaling}\label{app:diffMI}
Here we show that the diffusion equation starting from pure noise leads to self-similar behaviour for the correlation function. Let us use $u(x,t)$ as our field to avoid confusion with the samples $\phi(x,t)$ in the main body of the paper.

We wish to solve 
\begin{flalign}
\frac{\partial u}{\partial t} &= \nabla^2 u \nonumber \\
u(x,0) &\sim \mathcal{N}(0,1)
\end{flalign}
where the initial condition on the second line indicates that each pixel in the starting pure noise sample is an independent identically distributed standard normal random variable. Note firstly that if we take the expectation of this equation, we see that we start at $\langle u(x,t) \rangle = 0$ and remain there for all time.

Since $u(x,t)$ itself is ultimately random, we want to instead study its statistics, which are deterministic; namely, we want to study its two-point correlation function
\begin{equation}
G_t(x,y) := \langle u(x,t)u(y,t)\rangle 
\end{equation}
It's important to note that in this context, since the mean $\langle u(x,t) \rangle = 0$ for all $t$, this definition is in this case equivalent to the connected two-point function $\langle u(x,t)u(y,t)\rangle - \langle u(x,t)\rangle \langle u(y,t)\rangle$.

Due to the translational invariance of the grid and the problem, $G_t(x,y)$ must also be translationally equivariant, meaning that it cannot depend on $x,y$ in particular, but only their separation:
\begin{equation}\label{eq:translation_equivariance_two_point_fn}
G_t(x,y) = f(x-y,t)
\end{equation}
For some function $f$. Let us therefore define the variable
\begin{equation}
z := x-y
\end{equation}

Now, we can multiply the $x$-evolution equation by $u(y,t)$ and then take the expectation value with respect to the random initial starting grid. This will give us
\begin{flalign}
\left\langle\frac{\partial u(x,t)}{\partial t}u(y,t)\right\rangle &= \langle u(y,t)\nabla_x^2 u(x,t) \rangle \nonumber 
\end{flalign}
Now we use the translation equivariance formulated in \eqref{eq:translation_equivariance_two_point_fn} to realise that on the right hand side, in this case, the derivatives wrt $x$ may be replaced by derivatives in $z$. therefore, the entire equation is
\begin{equation}
\frac{1}{2}\frac{\partial G(z,t)}{\partial t} = \nabla_z^2 G(z,t)
\end{equation}
We get the factor of $1/2$ on the left hand side since there is another term that comes from the time derivative acting on $u(y,t)$ from the chain rule, which is however identical to the term we already have because of translation equivariance.

Therefore now we want to solve another diffusion equation for the two-point correlation, but now with a deterministic initial condition:
\begin{equation}
G(z,0)=G(x-y,0) = \langle u(x,0) u(y,0) \rangle = \delta(x-y) = \delta(z).
\end{equation}
Since the pixels were i.i.d. normally distributed.

The easiest way to solve this equation explicitly is a Fourier transform in space, wherein
\begin{flalign}
\frac{1}{2} \frac{\partial \tilde{G}(k,t)}{\partial t} &= -k^2\tilde{G}(k,t) \nonumber \\ \tilde{G}(k,0) &= 1,
\end{flalign}
which is straightforward to solve, giving
\begin{equation}
\tilde{G}(k,t) = e^{-2k^2 t} \ . 
\end{equation}
Fourier transforming back to position space finally yields
\begin{equation}
G(z,t) = \sqrt{\frac{\pi}{2t}}e^{-\pi^2 z^2/2t}.
\end{equation}
In general, the important feature of all such solutions is the distinctive behaviour
\begin{equation}
G(z,t) = g(t)h(z^2/t)
\end{equation}
where all spatial behaviour is controlled by the \textit{self-similar} variable $z^2/t$. This is a consequence of the fact that in the original diffusion equation, $z$ and $t$ scale in such a way that $z^2/t$ is a dimensionless variable.

\section{Experimental details}

In all our experiments, we train a diffusion on model on 2D image data, using a convolutional neural network architecture (details for each experiment below). We use the standard DDPM forward process, and the ODE formulation of the reverse process as laid out in \eqref{eq:ODE_diff}, with a simple first-order Euler integrator.

\subsection{Diffusion model trained on $\pm 1$ grids}
Here, we train a diffusion model on two $64 \times 64$ grids, one consisting of $+1$ at every pixel, and the other of $-1$ at every pixel. Due to the highly symmetrical, zero-average nature of the dataset, numerical instability and poor training are easily encountered. To fix this, we trained using $v$-prediction \cite{Salimans:2022} rather than prediction of the score function $\pi$ directly.

$v$-prediction works by learning the velocity vector at each moment of the reverse process rather than the score function:
\begin{equation}
v_t = \sqrt{\bar{\alpha}_t} \epsilon -\sqrt{1-\bar{\alpha}_t}\phi_0
\end{equation}
Where $\phi_0$ are the initial samples.

It is also possible to immediately obtain $\epsilon$ from learning $v$, but in practice learning $v$ is more stable. We use a forward process with a finishing time of $T=1000$, with a batch size of $32$ with the two samples listed above. The noise is added according to the standard cosine scheduling in \cite{nichol2021improved}:
\begin{equation}
\alpha_t = \cos^2 \left(\frac{\pi}{2} \frac{\frac{t}{T}+s}{1+s}\right)
\end{equation}
We amortise over time, but over an interval $(0,0.8T)$, since we deliberately avoid the very end of the forward process, where we are near the pure noise endpoint, and $\epsilon$ carries little data, but is instead predominantly noisy. That is, the loss function of training is
\begin{equation}
\mathcal{L}_v = \mathbb{E}_{t\in(0,0.8T), \epsilon \sim \mathcal{N}(0,I),\phi_0\sim data} \left[ |v_\theta(\phi_t,t) -v_t|^2\right]
\end{equation}
where $\theta$ are the neural network parameters.

Training is carried out over $200$ steps, using an ADAM optimiser with learning rate $= 2 \times 10^{-4}$.

The network has a convolutional $\mathbb{Z}_2$-equivariant local architecture, with $4$ $2D$ convolutional layers, each with a kernel size of $3$, and the number of channels following $(1,2,2,1)$ for this simple task with two binary homogenous samples. Time is embedded additively after the first Conv2D layer, via a simple MLP with $2$ linear layers sandwiching a SiLU activation.

Once $v$ is learned, we recover $\epsilon, \phi_0$ from its definition, and use these to implement the Euler integrator reverse process as usual. The score function used in the reverse process is, as per usual score matching,
\begin{equation}
s_\theta(\phi_t,t)=-\frac{1}{\sqrt{1-\bar{\alpha}_t}} \epsilon_\theta(\phi_t,t)
\end{equation}

The reverse process is implemented with $400$ steps and an output batch size of $20$, from which statistics on correlation are extracted.

\subsection{Diffusion model trained on MNIST}

Next, we train the diffusion model on MNIST, a dataset of $28 \times 28$ black-and-white images of digits $0$ to $9$. Here, we again use a final time in the forward process of $T=1000$. 

For this more complicated task, we use a more standard UNet architecture, however \textit{without} ConvTranspose2D layers, since these break locality and diverge from the theoretical ELS machine predictions. We use a minimal architecture with two downsampling blocks (going up to 128 channels), and two upsampling blocks, with no skip connections. 

In this case, the dataset is not quite so pathological as in the previous example, and $v$-prediction is not required. It is simpler to just train the network to learn $\epsilon$, with the standard loss function:
\begin{equation}
\mathcal{L} = \mathbb{E}_{t\in(0,0.8T), \epsilon \sim \mathcal{N}(0,I),\phi_0\sim data} \left[ |\epsilon_\theta(\phi_t,t) -\epsilon_t|^2\right]
\end{equation}
The noise is also simpler in this implementation: we use a linear schedule where $\beta_t$ runs linearly across the interval $(0.0001, 0.02)$, and then $\bar{\alpha}_t = \prod_{i=0}^t (1-\beta_t)$ as usual. The cosine schedule is not needed. 

We then use the learned $\epsilon$ as usual to drive the reverse process, which is once more a first-order Euler integrator, this time with 200 steps, and an output batch size of 100.

\subsection{ELS implementation}
In several cases, we implement the ELS machine exactly to compare against the results of our local models. This serves two purposes: firstly, checking that models are behaving as expected, and also providing a first test that our theory of correlations works in an analytic setting where we have more control and knowledge, as opposed to a trained model.

For the $\pm 1$ grid example, the analytical description \eqref{eq:ELS_ones_minus_ones_evolution_equation} makes implementation of the reverse process exact and straightforward. For MNIST, local $5\times5$ patches (though the size can easily be adjusted) are first extracted using periodic boundary conditions (to keep the theoretical setting of the ELS as simple as possible) from the dataset, before then being used in the weighted sum to calculate the score function. This is the most computationally intensive part of all the code that was run. After this, the score function is used as normal to implement the reverse process.

\subsection{Calculating correlations}
The correlation of two pixels \eqref{eq:rho_definition} is calculated as follows: for each grid, first the standard deviation is calculated across all pixels in the grid, and then the two-point function is calculated. The mean is taken at each separation $r$ of all the two point functions of all pixels separated by the vector $(r,0)$; that is, a distance $r$ along the $x$-axis. This mean is normalised by the standard deviation to obtain $\rho(r)$ for each grid sample, and finally the average is taken across the entire batch of samples to arrive at a final statistic. It is important to calculate $\rho$ per sample first, to ensure that the averages are being evaluated correctly.

As stated below the plots in the main body of the texts, the very smallest separations ($r\sim1,2$ pixels) are excluded from the plots as our theory is not meaningful at this scale (the effective smearing across a patch, as detailed int the text, now contains the lengthscale we want to examine). However, we also only take $r < L/2$, with $L$ being the total size of the grid, since beyond this scale, periodicity affects the correlation function and makes it increase again after decaying with $r$. This is outside the scope of the behaviour we study.

\end{document}